\documentclass[reprint,onecolumn,
amsmath,amssymb,aps]{revtex4}
\usepackage{epsfig,bm,epsf,graphics}
\usepackage{graphicx}
\usepackage{dcolumn}
\usepackage{bm}
\usepackage{xcolor}
\begin{document}
\draft
\preprint{APS/123-QED}
\title{Frustrated Antiferromagnetic Triangular Lattice with
Dzyaloshinskii-Moriya Interaction: Ground States, Spin Waves, Skyrmion Crystal, Phase Transition}
\author{Sahbi El Hog $^1$\footnote{haksahbi@gmail.com},  Ildus F. Sharafullin $^2$\footnote{sharafullinif@yandex.ru},
H. T. Diep $^{3}$\footnote{diep@cyu.fr, corresponding author},   H. Garbouj $^1$\footnote{garboujhedi3@gmail.com},
M. Debbichi$^1$\footnote{mourad\_fsm@yahoo.fr} and M. Said $^1$\footnote{moncef\_said@yahoo.fr }}
\affiliation{%
$^1$ Laboratoire de la Mati\`ere Condens\'ee \& Nanosciences
Facult\'e des Sciences de Monastir, Monastir, Tunisia\\
$^{2}$ Institute of Physics and Technology, Bashkir State University, 32, Validy str, 450076 Ufa, Russia\\
$^3$ Laboratoire de Physique Th\'eorique et Mod\'elisation,
CY Cergy Paris Universit\'e, CNRS, UMR 8089\\
2, Avenue Adolphe Chauvin, 95302 Cergy-Pontoise Cedex, France.\\
 }%

\date{\today}

\begin{abstract}
We study in this article a triangular lattice with Heisenberg spins interacting with each other via an antiferromagnetic exchange interaction $J$ and a Dzyaloshinskii-Moriya (DM) interaction $D$, between nearest-neighbors (NN).  We consider two cases: the first case in which the DM vector $\mathbf D$ is perpendicular to the lattice plane and the second case where it lies in the plane. A magnetic field $H$ is applied perpendicular to the spin plane in both cases. The ground state (GS) of this system is calculated by minimizing the energy using the very fast steepest-descent method in the two cases.  In the case of perpendicular $\mathbf D$ with $H=0$, the GS is periodic. We analytically determine the GS configuration which is characterized by two well-defined angles. We calculate the spin-wave spectrum in this case which shows that for small wave vectors the spin waves are forbidden in the system. When $H\neq 0$, the GS in the perpendicular $\mathbf D$ case shows no skyrmions. However, in the case of in-plane $\mathbf D$ with $H\neq 0$, we find a crystal of skyrmions at $T=0$ composed of three interpenetrating skyrmion sublattice crystals, in agreement with low-$T$ spin textures found in earlier works.
We show by Monte Carlo simulations that this skyrmion crystal is stable at finite temperatures below a critical temperature.
\vspace{0.5cm}
\begin{description}
\item PACS numbers: 75.25.-j ; 75.30.Ds ; 75.70.-i \\
\item Keywords: Frustrated antiferromagnetic triangular lattice;
Dzyaloshinskii-Moriya interaction; spin waves; skyrmions;
phase transition; Green's function theory; Monte Carlo simulation.
\end{description}
\end{abstract}


\maketitle

\section{Introduction}

Theoretical and experimental investigations on the effect of the Dzyaloshinskii-Moriya (DM) interaction in various materials have been extensively carried out in the context of weak ferromagnetism observed in perovskite compounds (see references cited in Refs. \cite{Sergienko,Ederer}). However, the interest in the DM interaction goes beyond the weak ferromagnetism. It has been shown that the DM interaction is at the origin of topological skyrmions \cite{Muhlbauer,Yu2,Yu1,Maleyev,Lin,Bogdanov,Rossler,Seki,Leonov2015,Adams,Heurich,Wessely,Jonietz} and new kinds of magnetic domain walls \cite{Heide,Rohart}.  The increasing interest in skyrmions results from the fact that skyrmions may play an important role in technological application devices \cite{Fert2013,Zhang}.
We note however that skyrmions can be created without the DM interaction provided that competing interactions between far neighbors have to be included to make the system over-frustrated \cite{zhang2017skyrmion,Zhang2020,Zhang2021,Okubo}.  The frustration is the origin of many new phenomena (see reviews in various domains given in Ref. \cite{DiepFSS}).

It is interesting to note that skyrmion crystals have been experimentally observed in various materials \cite{Yu2,Gilbert,Muhlbauer,Yu3}, but the most beautiful skyrmion crystal was observed in two-dimensional Fe(0.5)Co(0.5)Si by Yu {\it et al.} using Lorentz transmission electron microscopy \cite{Yu1}.

There has been also an increasing number of computational investigations of the spin texture by micromagnetism calculations \cite{Zhang,zhang2017skyrmion,Zhang2020,Zhang2021} and the steepest-descent method \cite{ElHog2018,sharafullin2019dzyaloshinskii,Sharafullin2020} amongst others \cite{Hayami,Okubo}. The spin texture called skyrmion is a vortex-like structure with Bloch-type or N\'eel-type. Their diameter is about ten lattice spacings, dependent of the applied field and other parameters.  We have shown that skyrmions originated from the DM interaction in competition with the ferromagnetic exchange interactions
form a superstructure on the underlying lattice: this superstructure is called "skyrmion crystal" which has been shown to be stable at finite temperatures \cite{ElHog2018,sharafullin2019dzyaloshinskii,Sharafullin2020}.

In this paper, we investigate the possibility of an existence of such skyrmion crystals on the frustrated triangular lattice taking into account the DM interaction in addition to the frustrating antiferromagnetic interaction $J$ between nearest-neighbors (NN) and an applied magnetic field $H$ perpendicular to the lattice plane. Note that the triangular antiferromagnet without DM interaction but with interaction between further neighbors has been shown to exhibit also a skyrmion crystal \cite{Hayami,Okubo}. When the DM vector $\mathbf D$ lies in the lattice plane and in a range of a magnetic field, we show in this paper that our model presents a skyrmion crystal composed of three interpenetrating skyrmion crystals resulting from the NN triangular antiferromagnet over-frustrated by the DM interaction. This model has been studied earlier by Rosales et al \cite{Rosales} using Monte Carlo (MC) simulations to cool the system from high temperatures to determine the spin configurations at low but finite temperatures. A quantum version of the model has been studied by Liu et al \cite{Liu} using a mean-field MC simulation to get low-temperature (low-$T$) skyrmion structures.
Very recently, a model including a single-ion anisotropy has been investigated by Mohylna and Zukovic \cite{Mohylna} after their earlier works on the same model \cite{Mohylna1,Mohylna2}.   Again here,  MC simulations have been used to determine the spin structure at low $T$.  We should also mention the work by Osorio et al. \cite{Osorio} on the same model using a mean-field approximation.
In our present work, we determine the skyrmion structure {\it at zero temperature}. We will see that our results are in agreement with the low-$T$ structures found in the above-mentioned works.

We also investigate in the present paper the case of perpendicular $\mathbf D$. We determine analytically the ground state (GS) and calculate the spin-wave spectrum and the local magnetization as a  function of temperature ($T$).  There is no skyrmion found in this case under any applied magnetic field.

Section \ref{DMI} recalls the DM interaction. Section \ref{Model} describes our model. In Section \ref{GSH0} we calculate analytically the GS of the spins on the triangular lattice interacting with each other via only a DM interaction with a perpendicular DM vector $\mathbf D$. Section \ref{SWT} shows the GS in the presence of both a perpendicular DM vector and an antiferromagnetic interaction $J$, in zero field. We present also in this section the calculation of the spin-wave spectrum and the magnetization up to the transition temperature. Section \ref{SC} shows the GS in the case of an in-plane DM vector in nonzero field where we find a skyrmion crystal composed of three skyrmion sublattices.  We present here the Monte Carlo (MC) results on the temperature dependence of physical quantities of the skyrmion crystal which indicate the stability of this phase up to a transition temperature. Concluding remarks are given in Section \ref{Concl}.

\section{Dzyaloshinskii-Moriya interaction}\label{DMI}
The DM interaction energy between two spins $\mathbf S_i$ and $\mathbf S_j$ is written as
\begin{equation}\label{eq1}
{\cal H_{DM}}=- \mathbf D_{i,j}\cdot \mathbf S_i\times \mathbf S_j
\end{equation}
where $\mathbf D_{i,j}$ is called "DM vector" which results from the displacement of non magnetic ions located in the bisecting plane between $\mathbf S_i$ and $\mathbf S_j$,  such as in Mn-O-Mn bonds considered in the historical papers \cite{Dzyaloshinskii,Moriya}. The direction of  $\mathbf D_{i,j}$ depends on the symmetry of the displacement \cite{Moriya}.  For two spins, the vector product in the DM interaction is antisymmetric with respect to the inversion symmetry. However, if $\mathbf D_{i,j}$ is antisymmetric with respect to the permutation of $i$ and $j$, then ${\cal H_{DM}}$ is symmetric. This point is important, this yields a non-zero energy when one sums the centro-symmetric NN spin pairs.

In the following we will choose  $|\mathbf D_{i,j}|=D$ independent of $(i,j)$ for simplicity. We note that $D$ can be assumed to be large, as it has been observed  experimentally. Let us mention an experimental value of a strong magnetoelectric coupling. The authors of Ref. \cite{Ferriani} found for a monolayer of Mn/W(110) a nearest-neighbor DM interaction with coupling constant $D$= 4.6 meV that dominates in the range where the exchange energy is small and leads to a large energy gain of about 6.3 meV/Mn atom (see their Fig. 2). Note that the first principle calculations \cite{Bode} give the exchange magnetic interaction of this system $J$=1.4 meV, and the anisotropy $K$=0.6 meV. The authors give the condition $D^2>4JK$ for spiral configurations. This means that $D$ should be larger than 1.83 meV.
In other magnetic materials, magnetic exchange interactions are of the order of a few dozens Kelvin: For MnTe $J$=20 K (2 meV) (see Refs. \cite{Szuszkiewicz,Magnin}), for perovskites LaSrMn3+Mn4+TiO, $J$=25 K (2.5 meV) (see Ref. \cite{Yahyaoui}). So, the DM coupling constant experimentally observed $D= 4.6$ meV is of the same order of magnitude as typical exchange interactions (even twice larger). This justifies the use of large $D$ in our model below.

\section{Model}\label{Model}
\begin{figure}[h!]
\centering
\includegraphics[width=6cm]{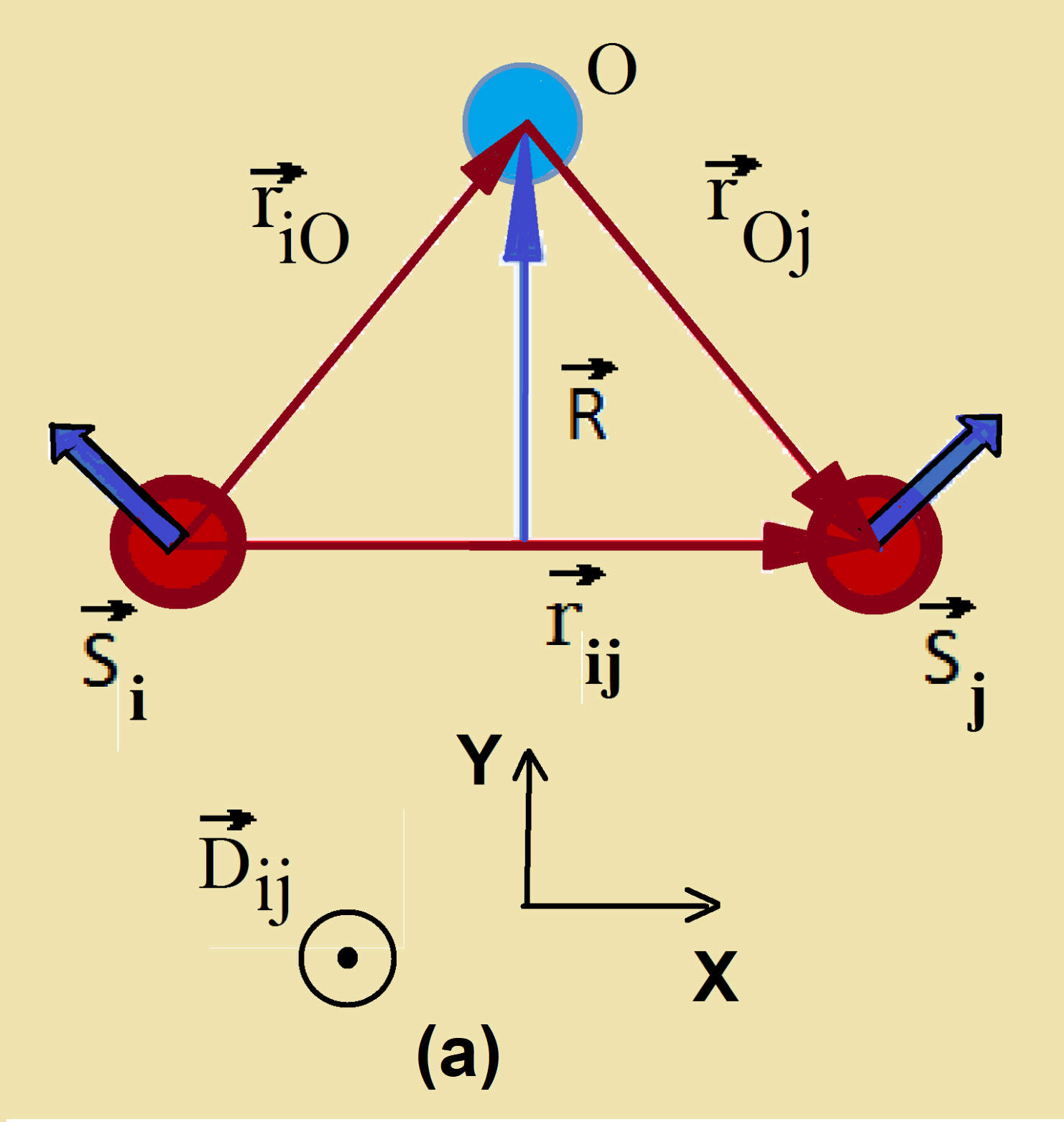}
\includegraphics[width=6cm]{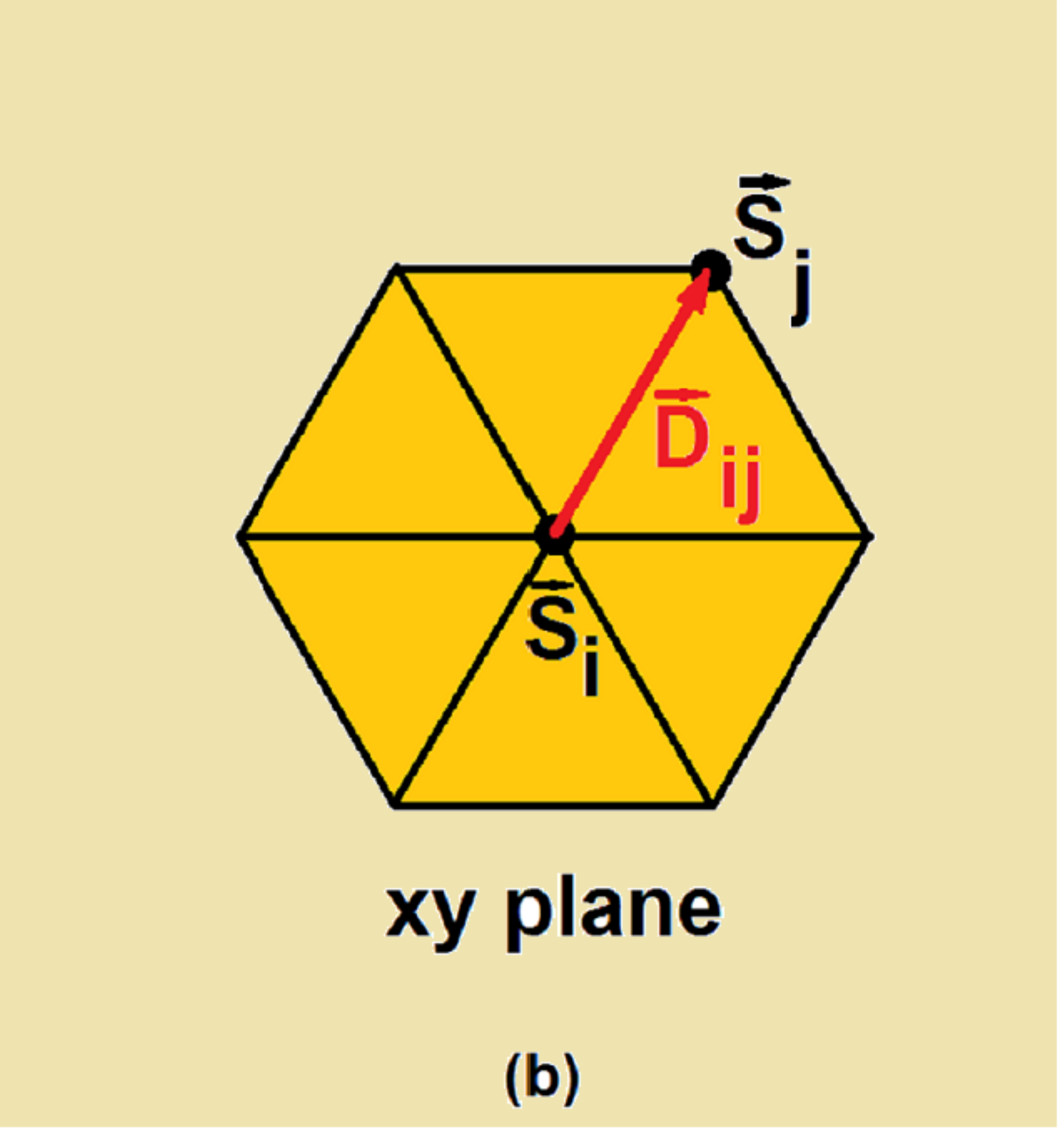}
\caption{(a)  $D$ vector along the $z$ direction perpendicular to the $xy$ plane. See the definition of the $D$ vector in the text, (b) In-plane $\mathbf D_{ij}$ vector chosen along the direction connecting spin $\mathbf S_i$ to spin $\mathbf S_j$ in the $xy$ plane.}\label{D}
\end{figure}

We consider a triangular lattice where the lattice site $i$ is occupied by a Heisenberg spin $\mathbf S_i$ of magnitude 1. We suppose that $\mathbf D_{i,j}$ is a vector perpendicular to the $xy$ plane and is given by \cite{Keffer,Cheong}
\begin{equation}\label{D1}
\mathbf D_{i,j}  \propto  \mathbf r_{ij} \times \mathbf R=- \mathbf r_{iO} \times \mathbf r_{Oj} 
\end{equation}
where $\mathbf r_{iO}=\mathbf r_O-\mathbf r_i$ and $\mathbf r_{Oj}=\mathbf r_j-\mathbf r_O$, $\mathbf r_{ij}=\mathbf r_j-\mathbf r_i$.   $\mathbf r_O$ is the position of non-magnetic ion (oxygen) chosen inside the triangle and $\mathbf r_i$ the position of the spin $\mathbf S_i$ etc. These vectors are defined in Fig. \ref{D}a in the particular case where the displacements  are in the $xy$ plane. We have therefore $\mathbf D_{i,j}$ perpendicular to the $xy$ plane in this case.

Note however  that if the atom displacements are in 3D space, $\mathbf D_{i,j}$ can be in any direction. In this paper, we consider also the case where $\mathbf D_{i,j}$ lies in the $xy$ plane as shown in Fig. \ref{D}b where $\mathbf D_{i,j}$ is taken along the vector connecting spin $\mathbf S_i$ to spin $\mathbf S_{j}$.

Note that from Eq. (\ref{D1}) one has
\begin{equation}\label{D2}
\mathbf D_{j,i}=-\mathbf D_{i,j}
\end{equation}

The DM interaction presented above was  historically introduced to explain the weak ferromagnetism in compounds MnO. The~superexchange between two Mn atoms is modified with the displacement of the oxygen atom between them. If~the displacement of the oxygen is in the $xy$ plane (see Figure~1a), then the DM vector $\mathbf D_{i,j}$ is perpendicular to the $xy$ plane and given by Eq. (1), taken from the historical Refs. \cite{Keffer,Cheong}.

Note however that the DM interaction goes beyond the context of weak ferromagnetism and may find its origin in various physical mechanisms. So, we define $\mathbf D_{i,j}$  in our work by Eq. (4) which respects the historical symmetry requirement Eq. (3). Our model (4) has nothing to do with the oxygen in compounds MnO. Our definition given in (4) is thus a model. The same for the case of parallel $\mathbf D_{i,j}$ given by Eq. (6) used by many authors.  There exist many other models for $\mathbf D_{i,j}$ used for example at the magneto-electric interface in ferroelectric multilayers.

It is clear that the definition of $\mathbf D_{i,j }$ (perpendicular or parallel) has nothing to do with the weak ferromagnetism, namely with MnO compounds. We recall that in the papers of Moriya and Dzyaloshinskii \cite{Moriya,Dzyaloshinskii}, the DM interaction  comes from the second-order perturbation of the exchange interaction $J$ between Mn NN atoms. Its coefficient $D$ is of the order of a few percents of $J$. This is correct for the very weak ferromagnetism observed in antiferromagnet MnO compounds.  At present, the DM interaction does not limit to the  second-order perturbation of the exchange interaction $J$, it has other physical origins and can be very large, of the order of $J$, this leads to the formation of skyrmions (see the discussion in the paragraph below Eq. (1)).

In the case of perpendicular $\mathbf D_{i,j}$, our model consists in
defining $\mathbf u_{i,j}$ as the unit vector on the $z$ axis. From Eqs. (\ref{D1})-(\ref{D2}) one writes
\begin{eqnarray}
\mathbf D_{i,j}&=&D\mathbf u_{i,j}\label{D3a}\\
\mathbf D_{j,i}&=&D\mathbf u_{j,i}=-D\mathbf u_{i,j}\label{D3b}
\end{eqnarray}
where $D$ represents the DM interaction strength.

In the case of in-plane $\mathbf D_{i,j}$, we define $\mathbf D_{i,j}$ as

\begin{equation}\label{PDM}
\mathbf D_{i,j}=D(\mathbf r_j-\mathbf r_i)/|\mathbf r_j-\mathbf r_i|=D\mathbf {r}_{ij}
\end{equation}
where $D$ is a constant and $\mathbf {r}_{ij}$ denotes the unit vector along $\mathbf r_j-\mathbf r_i$.

\section{Ground State with a Perpendicular $\mathbf D$ in Zero Field}\label{GSH0}

The Hamiltonian is given by

\begin{eqnarray}
\mathcal{H}&=&-J \sum_{\langle ij \rangle} \mathbf {S_i} \cdot \mathbf{S_j} -
 D \sum_{\langle ij \rangle} \mathbf u_{i,j} \cdot \mathbf {S_i} \times \mathbf {S}_{j} \nonumber\\
&&-H \sum_i S_i^z
\end{eqnarray}
where $\mathbf {S_i}$ is a classical Heisenberg spin of magnitude 1 occupying the lattice site $i$. The first sum runs over all spin nearest-neighbor (NN) pairs  with an antiferromagnetic exchange interaction $J$ ($J<0$), while the second sum is performed over all  DM interactions between NN.  $H$ is the magnitude of a magnetic field applied along the $z$ direction perpendicular to the lattice $xy$ plane.

In the general case, to determine the ground state (GS), we use the so-called "steepest descent method" which consists in minimizing the energy of each spin, one after another for the whole  system. We say we make one iteration.
To minimize the energy of a spin $i$ interacting with its 6 neighbors $j$, we calculate its energy $\mathcal H_i$. If $i$ is surrounded by 6 NN numbered from 1 to 6, then
\begin{eqnarray}
\mathbf{H_i}&=&-J  [S_i^xS_1^x+S_i^yS_1^y+S_i^zS_1^z+...] \nonumber\\
&&- D [ S_i^xS_{1}^y-S_i^yS_1^x +...] - HS_i^z
\end{eqnarray}
where we have taken the case of perpendicular $\mathbf D_{i,j}$ for demonstration. The three points $...$ in the first brackets denotes the remaining NN of the exchange term, while the three points $...$ in the second brackets denotes  the remaining NN terms of the DM interaction. Care should be taken on the sign of opposite $u_{ij}$.
Factorizing the above equation with respect to $S_i^x$, $S_i^y$ and $S_i^z$, we have
\begin{eqnarray}
\mathbf{H_i}&=&-S_i^x\{+J [S_1^x +S_2^x+...] + D  [S_{1}^y +...]\}\nonumber\\
&&-S_i^y \{+J[ S_1^y+S_2^y+...] +D [-S_{1}^x +...]\}-S_i^z\{J[S_1^z+S_2^z+...]+H\}\nonumber\\
&=&-S_i^xH^x-S_i^yH^y-S_i^zH^z
\end{eqnarray}
where $H^x$, $H^y$ and $H^z$ are the quantities in the respective curly brackets. These quantities are components of the interaction field $\mathbf H$ acting on the spin $\mathbf S_i$ from its neighbors. We write
\begin{equation}
\mathbf{H_i}=-\mathbf S_i \cdot \mathbf H
\end{equation}
In the above equation the state of the spin $\mathbf S_i$, namely $S_i^x$, $S_i^y$ and $S_i^z$, is from its current state. To minimize its energy, we
choose its new direction $\mathbf S_i'$ along the field $\mathbf H$. In doing so, its energy is minimum. The algorithm is very simple, we take
\begin{eqnarray}
(S_i^x)'&=& H^x/\sqrt{(H^x)^2+(H^y)^2+(H^z)^2}\\
(S_i^y)'&=& H^y/\sqrt{(H^x)^2+(H^y)^2+(H^z)^2}\\
(S_i^z)'&=& H^z/\sqrt{(H^x)^2+(H^y)^2+(H^z)^2}
\end{eqnarray}
We see that $|\mathbf S_i'|=1$ and $\mathbf S_i'$ is parallel to $\mathbf H$, making its energy minimum.
Next we take another spin and minimize its energy in the same manner until all spins are considered. This is one iteration.
For a given initial spin configuration, we repeat a large number of iterations, many hundred times or even many thousand times until the total energy gets the minimum value. This is one run. Care should be taken to check the minimum value by taking several thousands of independent runs with different initial spin configurations. The algorithm is very fast, allowing such a large number of runs. In general, it is known that when the interactions are periodic, namely no bond disorder as in spin glasses, this method gives true GS spin configurations \cite{Diep2013,DiepTM}. The system is not trapped in  meta-stable states.  A meta-stable state is a state due to the fact that we minimize the the energy of a spin, one after another: a spin is put in its minimized energy state using the current state of its neighbors. However, if  one or several of its neighbors change the state after, then the previously considered spin is no more in its minimized energy state. This is called  a meta-stable state. But in the steepest-descent method, we repeat iteratively the minimization procedure untill all spins are in their minimized energy state.

We consider a triangular lattice of lateral dimension $L$. The total number of sites $N$ is given by $N=L \times L$.  Though we use the periodic boundary conditions, to avoid the finite size effect, we have to find the limit beyond which the GS does not depend on the lattice size.

In the absence of $D$ the antiferromagnetic interaction on a triangular lattice causes a 120-degree structure in the $xy$ plane \cite{DiepFSS}.

\begin{figure}[h!]
\centering
\includegraphics[width=7cm]{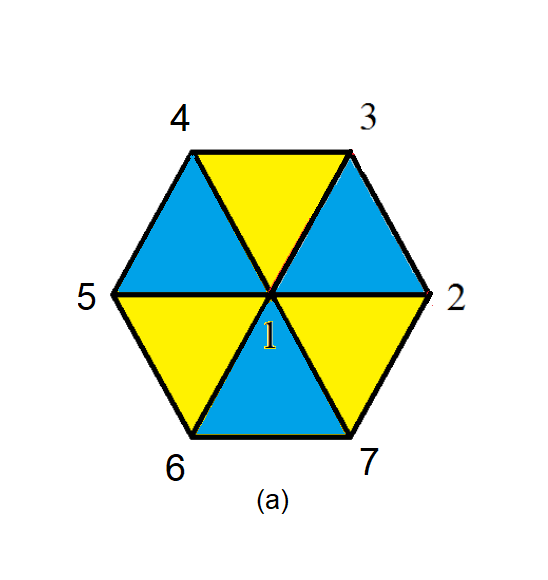}
\includegraphics[width=7cm]{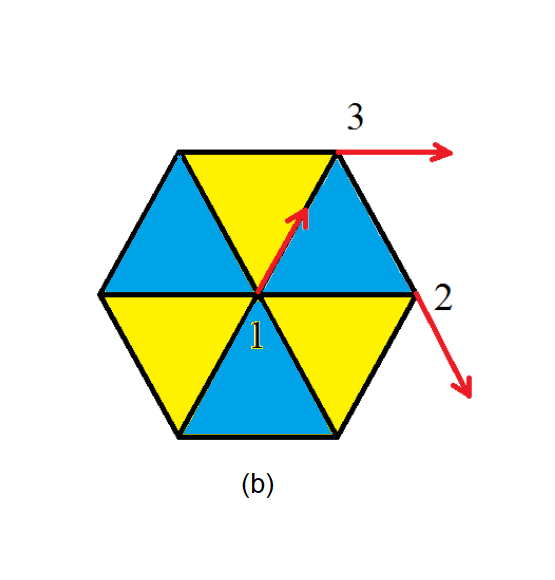}
\includegraphics[width=9cm]{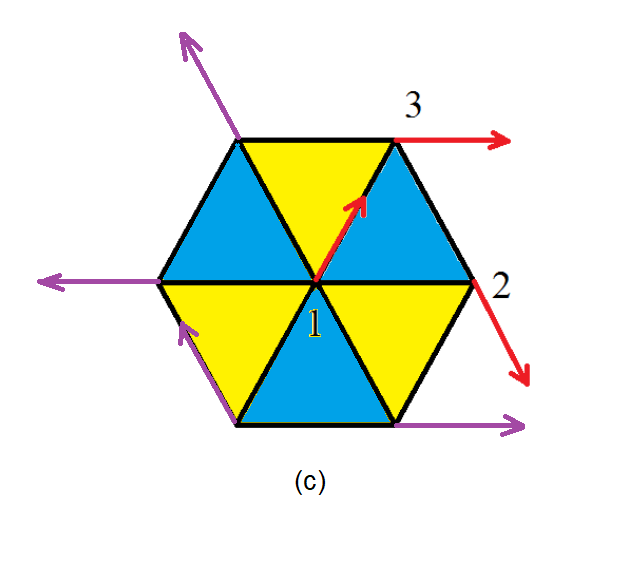}
\caption{Perpendicular $\mathbf D_{i,j}$: (a) The spins on a hexagon are numbered as indicated; (b)  Ground-state spin configuration with only Dzyaloshinskii-Moriya interaction on the triangular lattice ($J=0)$ is analytically determined. One angle is 120 degrees, the other two are 60 degrees. Note that the choice of the 120-degree angle  in this figure is along the horizontal spin pair. This configuration is one ground state, the other two ground states have the 120-degree angles on respectively the two diagonal spin pairs. Note also that the spin configuration is
invariant under the global spin rotation in the $xy$ plane; (c) example of construction of the GS for the whole hexagon: spins in violet are added in respecting the turning angle in each direction. See text for explanation.}\label{DMGSfig}
\end{figure}

In the absence of $J$, unlike the bipartite square lattice where one can arrange the NN spins
to be perpendicular with each order in the $xy$ plane, the triangular lattice cannot fully satisfy the
DM interaction for each bond, namely with the perpendicular spins at the ends.  For this particular case of interest, we can analytically calculate the GS spin configuration as shown in the following. One considers a hexagon with the spin at the center numbered 1 and the 6 spins around it numbered from 2 to 7 as indicated in Fig. \ref{DMGSfig}a.   
The DM energy of a plaquette is written as

\begin{eqnarray}
H_p&=& -D[\mathbf u_{1,2}\cdot \mathbf S_1 \times \mathbf S_2 +
       \mathbf u_{1,3}\cdot \mathbf S_1 \times \mathbf S_3 +
       \mathbf u_{1,4}\cdot \mathbf S_1 \times \mathbf S_4\nonumber\\ 
      &&+ \mathbf u_{1,5}\cdot \mathbf S_1 \times \mathbf S_5+
         \mathbf u_{1,6}\cdot \mathbf S_1 \times \mathbf S_6+
          \mathbf u_{1,7}\cdot \mathbf S_1 \times \mathbf S_7]
 \end{eqnarray}
Note  that the $u$ vectors of a pair of spins for example $\mathbf u_{1,2}$ is opposite of the pair on the opposite direction of (1,2), namely  $\mathbf u_{1,2}=-\mathbf u_{1,5}=1$. The same observation are for the other opposite pairs, namely  $\mathbf u_{1,3}=-\mathbf u_{1,6}=1$,  $\mathbf u_{1,4}=-\mathbf u_{1,7}=1$. In  addition the vector products of the opposite pairs are also opposite, for instance  $\mathbf S_1 \times \mathbf S_2=- \mathbf S_1 \times \mathbf S_5$ because of the chiral angles in the opposite directions. The same thing holds for the other opposite pairs.  This retains only 3 terms of the above equation with the coefficient 2:
\begin{eqnarray}
H_p &=& -2D[\sin \theta_{1,2}+\sin \theta_{1,3}+\sin \theta_{1,4}]\nonumber\\
&=& -2D[\sin \theta_{1,2}+\sin \theta_{1,3}+\sin \theta_{2,3}] \label{Hp}
\end{eqnarray}
Note that due to the symmetry we have replaced $\sin \theta_{1,4}$ by $\sin \theta_{2,3}$ sine they have the same chiral direction (see Fig. \ref{DMGSfig}a).
We recall that $\theta_{1,2}=\theta_2-\theta_1$ is the oriented angle between $\mathbf S_1$ and $\mathbf S_2$, etc.
The expression (\ref{Hp}) was written in the form (\ref{Hp}) so that it suffices to determine the relative angles between spins 1, 2 and 3 in a triangle in the GS. The GS of the whole system is constructed by using the chiral angle in each direction (1,2), (1,3) and (2,3).


The minimization
of $H_p$ yields
\begin{eqnarray}
\frac{dH_{p}}{d\theta_1}&=&0=
-2D[ -\cos (\theta_{2}-\theta_{1})-\cos (\theta_{3}-\theta_{1})]\label{minimize1}\\
\frac{dH_{p}}{d\theta_2}&=&0=  -2D[ \cos (\theta_{2}-\theta_{1})- \cos (\theta_{3}-\theta_{2})]\label{minimize2}\\
\frac{dH_{p}}{d\theta_3}&=&0=  -2D[\cos (\theta_{3}-\theta_{2}) +\cos (\theta_{3}-\theta_{1})]
\label{minimize3}
\end{eqnarray}
The solutions for the above equations are $\theta_{1,2}=\pm\theta_{1,3}$, $\theta_{2,3}=\pm\theta_{1,2}$ and
$\theta_{1,3}=\pm\theta_{2,3}$. We have to choose the correct sign in each spin pair  so as the relative angle between this  NN spin pair is not zero. Othewise, if the relative angle is zero,  the interaction energy of such a pair yields the zero DM energy. The correct choices of sign finally give
\begin{eqnarray}
\theta_{1,2}&=&-\theta_{1,3} \ \ \mbox{so that} \ \ \theta_{2,3}=\theta_{2,1}+\theta_{1,3}=-2\theta_{1,2}\label{gsangle1}\\
\theta_{2,3}&=&\theta_{1,2} \ \ \mbox{so that} \ \ \theta_{1,3}=\theta_{1,2}+\theta_{2,3}=2\theta_{2,3}\label{gsangle2}\\
\theta_{1,3}&=&-\theta_{2,3} \ \ \mbox{so that} \ \ \theta_{2,1}=\theta_{2,3}+\theta_{3,1}=-2\theta_{1,3}\label{gsangle3}
\end{eqnarray}
These three equations, Eqs. (\ref{gsangle1})-(\ref{gsangle3}), should be solved.  We have from Eq. (\ref{minimize1})
$\cos (\theta_{1,2})=-\cos (\theta_{1,3})$. Using Eq. (\ref{gsangle3}) one obtains
\begin{equation}
\cos (2\theta_{3,1})=-\cos (\theta_{3,1})\ \ \rightarrow 2\cos^2(\theta_{3,1})+\cos(\theta_{3,1})-1=0
\end{equation}

This second-degree equation gives $\cos(\theta_{3,1})=\frac{-1\pm \sqrt{1+8}}{4}$. Only the solution with plus sign is acceptable so that  $\theta_{3,1}=\theta_{2,3}=\pi/3$. From Eq. (\ref{gsangle3}), one has  $\theta_{2,1}=2\pi/3$.
This is one solution summarized by Eq. (\ref{sol1}) below.
Note that we have taken one of them, Eq. (\ref{gsangle3}), to obtain explicit solutions for the three angles given in Eq. (\ref{sol1}). We can do the same calculation starting with Eqs. (\ref{gsangle1})-(\ref{gsangle2}) to get explicit solutions given in Eqs. (\ref{sol2})-(\ref{sol3}). We note that when we make a circular permutation of the indices of Eq. (\ref{sol1}) we get Eq. (\ref{sol2}), and a circular permutation of Eq. (\ref{sol2}) gives Eq. (\ref{sol3}).
One summarizes the three degenerate solutions below
\begin{eqnarray}
\theta_{3,1}=\theta_{2,3}=\pi/3, \ \ \theta_{2,1}=2\pi/3\label{sol1}\\
\theta_{1,2}=\theta_{3,1}=\pi/3, \ \ \theta_{3,2}=2\pi/3\label{sol2}\\
\theta_{2,3}=\theta_{1,2}=\pi/3, \ \ \theta_{1,3}=2\pi/3\label{sol3}
\end{eqnarray}
We show in Fig. \ref{DMGSfig}b the spin orientations of the solution (\ref{sol1}) using the trigonometric angle sense. The ground state energy is obtained by replacing the angles into Eq. (\ref{Hp}). For the three solutions, one gets the following energy of the hexagon
\begin{equation}\label{GSE}
H_p= -3D\sqrt{3}
\end{equation}
We have three degenerate ground states, in addition to the global arbitrary rotation of all spins in the $xy$ plane. Note that we plot the solution (\ref{sol1}) in  Fig. \ref{DMGSfig}b by choosing a particular spin configuration (spins on chiral axes) for an easy visualization, but any arbitrary global rotation is a solution. 

To construct the GS of the whole system, one starts from the triangle (1,2,3) of Fig. \ref{DMGSfig}b
and adds the next spin on each axis (1,2), (1,3) and (2,3) with the turning angle on each axis:  the horizontal axis (1,2) and those parallel to it have the turning angle of 2$\pi$/3, the axis (1,3) and (2,3) have the turning angle of $\pi/3$.  We should respect the sign of the turning angle while going to the next spin. An example of construction from the red spins of the triangle (1,2,3) is shown in Fig. \ref{DMGSfig}c for the hexagon: the added spins are violet. By adding spins, one by one in all directions, we generate the GS of the whole system. 

\begin{figure}[h!]
\centering
\vspace{1cm}
\includegraphics[width=12cm]{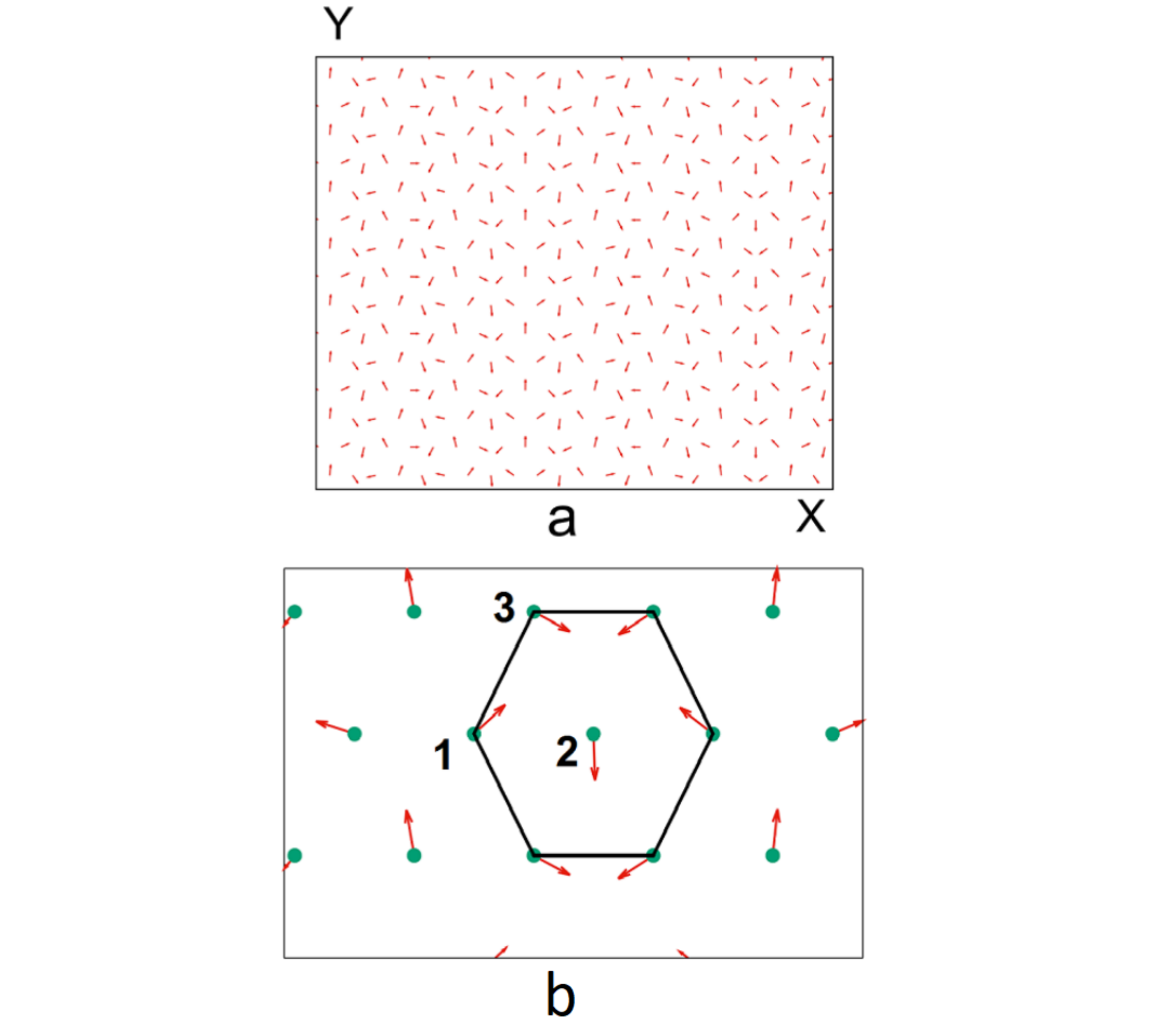}
\caption{Perpendicular $\mathbf D_{i,j}$: (a) Ground-state spin configuration with only Dzyaloshinskii-Moriya interaction on the triangular lattice ($J=0)$ obtained numerically by the steepest descent method, (b)  a zoom on a hexagonal cell, this is exactly what obtained analytically shown in Fig. \ref{DMGSfig} with a global spin rotation in the $xy$ plane: the angle of the horizontal pair (1,2) is 120 degrees, those of (2,3) and (3,1) are equal to 60 degrees.}\label{DMGSLfig}
\end{figure}

This solution can be numerically obtained by
the steepest descent method described above. The result is shown in Fig. \ref{DMGSLfig} for the full lattice.
We see in the zoom that the spin configuration on a plaquette is what obtained analytically, with a global spin rotation as explained in the caption of Fig.  \ref{DMGSfig}.

As said above, to use the steepest-descent method, we consider a triangular lattice of lateral dimension $L$. The total number of sites $N$ is given by $N=L \times L$. To avoid the finite size effect, we have to find the size limit beyond which the GS does not depend on the lattice size.  This is found for $L\geq 100$. Most of calculations have been performed for $L=100$, although for the phase transition we have checked that the transition temperature did not significantly depend on $L$ up to $L=200$.

\section{Ground State with both perpendicular $\mathbf D$ and $J$ in Zero Field- Spin Waves}\label{SWT}

\begin{figure}[h!]
\center
\includegraphics[width= 12cm]{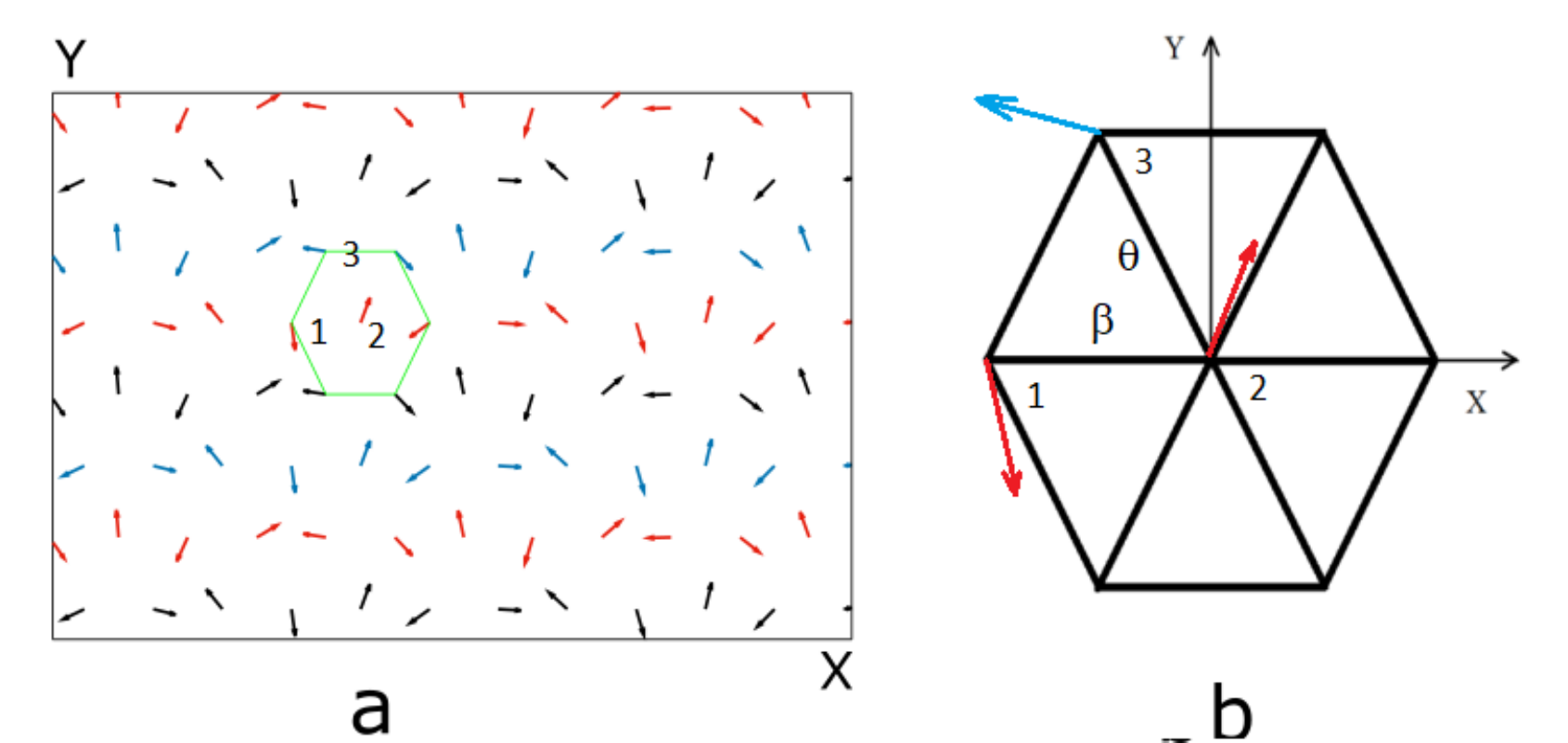}
\caption{Perpendicular $\mathbf D_{i,j}$ with antiferromagnetic $J$: (a) Ground-state spin configuration in zero field for $D = 0.5$, $J = -1$ where the angles in a hexagon are shown in (b) with $\beta = 156$ degrees for the pair (1,2) on the horizontal axis and  $\theta =  102$ degrees for the pairs (2,3) and (3,1) on the diagonals. Note that there are two other degenerate states where $\beta$ is chosen for the pair (2,3) or (3,1). }.\label{GSDJH0}
\end{figure}

When both $J$ and perpendicular $\mathbf D$ are present, a compromise is established between these competing interactions.  In zero field, the GS shows non-collinear but periodic in-plane spin configurations.  The planar spin configuration is easily understood: when $\mathbf D$ is perpendicular and without $J$, the spins are in the plane. When $J$ is antiferromagnetic without $\mathbf D$, the spins are also in the plane and form a 120-degree structure. When $\mathbf D$ and $J$ exist together the angles between NN's change but they still in the plane  in order to keep both $D$ and $J$ interactions as low as possible. An example is shown in Fig. \ref{GSDJH0} where one sees that the GS is planar and characterized by two angles  $\theta=102 $ degrees and one angle  $\beta=156$ degrees formed by three spins on a triangle plaquette.  Note that there are three degenerate states where $\beta$ is chosen for the pair (1,2) (Fig.   \ref{GSDJH0}a) or the pair (2,3) or the pair (3,1). Changing the value of $D$ will change the angle values. Changing the sign of $D$ results in a change of the sense of the chirality, but not the angle values.

In the case of perpendicular $\mathbf D_{i,j}$ in zero-field, as shown above we find the GS on a hexagon of the lattice  is defined by four identical angles
$\beta$ and two angles $\theta$ as shown in Fig. \ref{GSDJH0}. The values of $\beta$ and $\theta$ depend on the value of $D$. We take $J=-1$ (antiferromagnetic) hereafter. For $D=0.5$ we have $\beta =  156$ degrees and $\theta = 102$ degrees. For $D=0.4$ we obtain $\beta =  108$ degrees and $\theta = 144$ degrees, using $N=60\times 60$.

The periodicity of the GS allows us to calculate the spin-wave spectrum in the following.

The model for the calculation of the spin-wave spectrum uses quantum Heisenberg spins of magnitude $1/2$, it is given by
\begin{eqnarray}
\mathcal H=-J\sum_{\left<i,j\right>}\mathbf{S_i}\cdot\mathbf{S_j}-D \sum_{\left<i,j\right>} \mathbf u_{i,j} \cdot \mathbf{S_i} \times \mathbf{S_j} - I \sum_{\left<i,j\right>} S_i^z~S_j^z \cos\theta_{ij}
\end{eqnarray}
where $\theta_{ij}$ is the angle between $\mathbf{S_i}$ and $\mathbf{S_j}$ and the last term is an extremely small anisotropy added to stabilize the spin waves when the wavelength $k$ tends to zero \cite{DiepTM,Mermin}. Note that, as seen in the definition of the DM interaction, in a centrosymetric system two opposite spin pairs have opposite $u_{ij}$ and opposite $\sin \theta_{ij}$.  For example on the $x$ axis we have $u_{ij}\sin \theta_{ij}$ (forward) and $u_{ij'}\sin \theta_{ij'}$ (backward). The latter is equal to $-u_{ij}\sin (-\theta_{ij})$ which is equal to $u_{ij}\sin \theta_{ij}$. Taking $u_{ij}=1$, we do not have $u_{ij}$ anymore and the sum on  $ j$ is taken forward, namely  $j>i$ on each axis to avoid the double sum on bonds.  This sum is indicated as $<i,j>$ in the above formula.

\begin{figure}[ht!]
\centering
\includegraphics[width=6cm,angle=0]{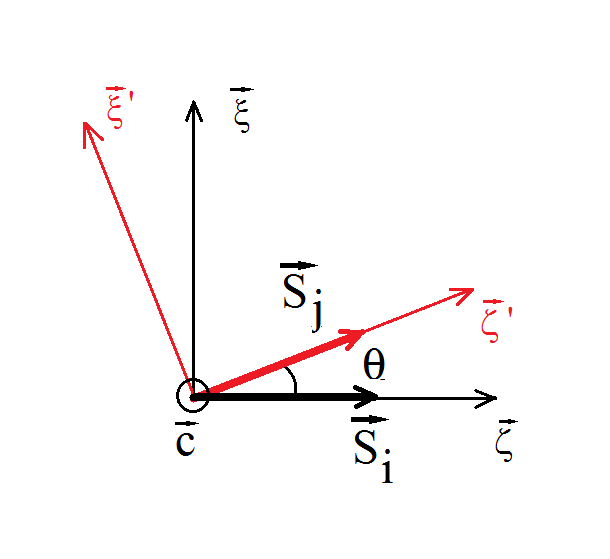}
\caption{Spin $\mathbf S_i$ lies along the $\vec \zeta$ axis (its quantization axis), while spin  $\mathbf S_j$ lies along its quantization axis $\vec \zeta'$ which makes an angle $\theta$ with the $\vec \zeta$ axis.  The axes $\vec \xi$ and $\vec \xi'$ are perpendicular respectively to $\vec \zeta$ and $\vec \zeta'$. The perpendicular axes $\hat \eta_i$ and $\hat \eta_j$ coincide with the $z$ axis, perpendicular to the $xy$ plane.}\label{localco}
\end{figure}

In order to calculate the spin-wave spectrum for systems of non-collinear spin configurations, let us emphasize that the commutation relations between spin operators are established when the spin lies on its quantization $z$.  In the non-collinear cases, each spin has its own quantization axis. It is therefore important to choose a quantization axis for each spin. We have to use the system of local coordinates defined as follows. In the Hamiltonian, the spins are coupled two by two. Consider a pair $\mathbf S_i$ and $\mathbf S_j$. As seen above,  in the general case these spins make an angle $\theta_{i,j}=\theta_j-\theta_i$ determined by the competing interactions in the systems. For quantum spins, in the course of calculation we need to use the commutation relations between the spin operators $S^z,S^+,S^-$. As said above, these commutation relations are derived from the assumption that the spin lies on  its quantization axis $z$.
We show in Fig. \ref{localco} the local coordinates assigned to spin $\mathbf S_i$ and $\mathbf S_j$. We write

\begin{eqnarray}
\mathbf S_i&=&S_i^x\hat \xi_i+S_i^y\hat \eta_i+S_i^z\hat \zeta_i\label{SI}\\
\mathbf S_j&=&S_j^x\hat \xi_j+S_j^y\hat \eta_j+S_j^z\hat \zeta_j\label{SJ}
\end{eqnarray}
Note that the spin components $S_i^x$ lies on $\hat \xi_i$, $S_i^y$  on $\hat \eta_i$ and $S_i^z$ on $\hat \zeta_i$. The indices $x$,$y$ and $z$ here have nothing to to with the real space orientations.  Expressing the axes of $\mathbf S_j$ in the frame of $\mathbf S_i$ one has
\begin{eqnarray}
\hat \zeta_j&=&\cos \theta_{i,j}\hat \zeta_i+\sin \theta_{i,j}\hat \xi_i\label{localco2}\\
\hat \xi_j&=&-\sin \theta_{i,j}\hat \zeta_i+\cos \theta_{i,j}\hat \xi_i\label{localco1}\\
\hat \eta_j&=&\hat \eta_i\label{localco3}
\end{eqnarray}
so that
\begin{eqnarray}
\mathbf S_{j}&=&S_j^x(-\sin\theta_{i,j}\hat \zeta_i +\cos \theta_{i,j}\hat \xi_i)\nonumber\\
&&+S_j^y\hat \eta_i+S_j^z(\cos\theta_{i,j}\hat \zeta_i +\sin \theta_{i,j}\hat \xi_i)\label{local6}
\end{eqnarray}

Using Eq. (\ref{local6}) to express $\mathbf S_j$ in the $(\hat \xi_i,\hat \eta_i,\hat \zeta_i)$ coordinates, then calculating
$\mathbf S_i \cdot \mathbf S_j$ and other terms of  Hamiltonian, we have
 (see Ref. \cite{Diep2017}):

\begin{eqnarray}
\mathcal {H}&=&-J\sum_{\left<i,j\right>} \frac{1}{4}(S_i^+S_j^++S_i^-S_j^-) (\cos\theta_{ij}-1)
+\frac{1}{4}(S_i^+S_j^-+S_i^-S_j^+) (\cos\theta_{ij}+1)\nonumber \\
&&+\frac{1}{2}S_j^z\sin\theta_{ij}(S_i^++S_i^-)-\frac{1}{2}\sin\theta_{ij}S_i^z(S_J^++S_j^-)+S_i^zS_j^z\cos\theta_{ij} \nonumber \\
&&-D\sum_{\left<i,j\right>} S_i^z S_j^z \sin\theta_{i,j} + \frac{1}{4}  \sin\theta_{i,j}( S_i ^+ S_j^+ + S_i ^+ S_j^- +
S_i ^- S_j^+)
+ \frac{1}{2}\cos\theta_{i,j}(S_i^z(S_j^+ + S_j^-) - S_j^z(S_i^+ + S_i^-))\nonumber \\
&&-I \sum_{\left<i,j\right>} S_i^z~S_j^z \cos\theta_{i,j} \label{generalH}
\end{eqnarray}

The Green's function theory for non-collinear spin configurations was first introduced in details 25 years ago \cite{Quartu1997}.  Following the method described in Ref. \cite{Diep2017} for the DM interaction, we define the Green's functions by
\begin{eqnarray}
G_{i,j}(t,t')&=&<<S_i^+(t);S_{j}^-(t')>>\nonumber\\
&=&-i\theta (t-t')
<\left[S_i^+(t),S_{j}^-(t')\right]> \label{green59a}\\
F_{i,j}(t,t')&=&<<S_i^-(t);S_{j}^-(t')>>\nonumber\\
&=&-i\theta (t-t')
<\left[S_i^-(t),S_{j}^-(t')\right]>\label{green60}
\end{eqnarray}
The equations of motion of these functions are written as

\begin{equation*}
\begin{aligned}
i\hbar \frac{dG_{i,j}(t-t')}{dt}&=2<S_i^z>\delta_{i,j}\delta(t-t')- J\sum_{\left<l\right>}  <S_i^z>F_{l,j}(t-t')(\cos\theta_{i,l}-1)\\
&+<S_i^z>G_{l,j}(t-t')(\cos\theta_{i,l}+1)-2\cos\theta_{i,l}<S_l^z>G_{i,j}(t-t')\\
&+D \sum_{\left<l\right>} 2\sin\theta_{i,l} <S_i^z>F_{l,j}(t-t') - \sin\theta_{i,l} <S_i^z>( G_{l,j}(t-t') + F_{l,j}(t-t')) \\
&- 2I \sum_{\left<l\right>} \cos\theta_{i,l} <S_i^z>F_{l,j}(t-t')
\end{aligned}
\end{equation*}

\begin{equation*}
\begin{aligned}
i\hbar \frac{dF_{i,j}(t-t')}{dt}&= J\sum_{\left<l\right>}  <S_i^z>G_{l,j}(t-t')(\cos\theta_{i,l}-1)\\
&+<S_i^z>F_{l,j}(t-t')(\cos\theta_{i,l}+1)-2\cos\theta_{i,l}<S_l^z>F_{i,j}(t-t')\\
&-D \sum_{\left<l\right>} 2\sin\theta_{i,l} <S_i^z>G_{l,j}(t-t') - \sin\theta_{i,l} <S_i^z>( G_{l,j}(t-t') + F_{l,j}(t-t')) \\
&+ 2I \sum_{\left<l\right>} \cos\theta_{i,l} <S_i^z>G_{l,j}(t-t')
\end{aligned}
\end{equation*}
Note that $<S_i^z>$ is the average of the spin $i$ on its local quantization axis in the local-coordinates system (see Ref. \cite{Diep2017}).
We use now the time Fourier transforms
of the $G$ and $F$ Green's functions, we get

\begin{equation}
\begin{aligned}
\hbar\omega g_{i,j}&=2\mu_{i} \delta_{i,j}- J\sum_{\left<l\right>}  \mu_{i} f_{l j}
e^{-i \mathbf k\cdot (\mathbf R_{i}- \mathbf R_{l})}     (\cos\theta_{i,l}-1)\\
&+\mu_{i} g_{l j} e^{-i \mathbf k \cdot (\mathbf R_{i}-\mathbf R_{l})} (\cos\theta_{i,l}+1)-2\mu_{l}\cos\theta_{i,l} g_{i,j}\\
&-D\sum_{\left<l\right>} 2\sin\theta_{i,l} \mu_{l} g_{i,j} - \sin\theta_{i,l} \mu_{i}( g_{l,j}e^{-i \mathbf k\cdot (\mathbf R_{i}-\mathbf R_{l})} +f_{l,j}e^{-i \mathbf k\cdot (\mathbf R_{i}-\mathbf R_{l})})\\
& + 2I \sum_{\left<l\right>} \mu_{l} \cos\theta_{i,l}  g_{i,j}\label{FourierG}
\end{aligned}
\end{equation}
and
\begin{equation}
\begin{aligned}
\hbar\omega f_{i,j}&= J\sum_{\left<l\right>}  \mu_{i} g_{l j}
e^{-i\mathbf k\cdot (\mathbf R_{i}-\mathbf R_{l})}(\cos\theta_{i,l}-1)\\
&+\mu_{i} f_{l j} e^{-i \mathbf k\cdot (\mathbf R_{i}-\mathbf R_{l})} (\cos\theta_{i,l}+1)-2\mu_{l}\cos\theta_{i,l} f_{i,j}\\
&+D\sum_{\left<l\right>} 2\sin\theta_{i,l} \mu_{l} f_{i,j} - \sin\theta_{i,l} \mu_{i}( g_{l,j}e^{-i \mathbf k\cdot(\mathbf R_{i}-\mathbf R_{l})} +f_{l,j}e^{-i \mathbf k\cdot (\mathbf R_{i}-\mathbf R_{l})})\\
& - 2I \sum_{\left<l\right>} \mu_{l} \cos\theta_{i,l}  f_{i,j}\label{FourierF}
\end{aligned}
\end{equation}
where $\mu_i\equiv <S^z_i>$,  $\mathbf k$ is the wave vector in the reciprocal lattice of the triangular lattice of the real space, and $\omega$ the spin-wave frequency.   Note that the index $z$ in $S^z_i$ is not referred to the real space direction $z$, but to the quantization axis of the spin $\mathbf S_i$.
At this stage, we have to replace $\theta_{i,j}$ by either $\beta$ or $\theta$ according on the GS spin configuration given above (see Fig. \ref{GSDJH0}).

Writing the above equations under a matrix form, we have
\begin{equation}
\mathbf M \left( \hbar\omega \right) \mathbf h = \mathbf C,
\label{eq:HGMatrix}
\end{equation}
where $\mathbf M\left(\hbar\omega\right)$ is a square matrix of dimension
$2\times 2$, $\mathbf h$ and $\mathbf C$ are
the column matrices which are defined as follows
\begin{equation}
\mathbf h = \left(%
\begin{array}{c}
  g_{i,j} \\
  f_{i,j} \\
\end{array}%
\right) , \hspace{1cm}\mathbf C =\left(%
\begin{array}{c}
  2 \left< S^z_i\right>\delta_{i,j}\\
  0 \\
\end{array}%
\right) , \label{eq:HGMatrixgu}
\end{equation}
and the matrix  $\mathbf M\left( \hbar\omega\right)$ is given by
\begin{equation*}
\hspace{-2cm}
\mathbf M \left( \hbar\omega\right)= \left(%
\begin{array}{*{2}c}
 \hbar\omega +A &B \\
   -B    & \hbar\omega-A \\
\end{array}%
\right)
\end{equation*}

The nontrivial solution of $g$ and $f$ imposes the following secular equation:

\begin{equation}\label{secular}
\hspace{-2cm}
0 = \left(%
\begin{array}{*{2}c}
  \hbar\omega+A &B \\
   -B    & \hbar\omega-A \\
\end{array}%
\right)
\end{equation}
where

\begin{equation}
\begin{aligned}
A&= -J(8 \mu_{i} \cos\beta (1+I) + 4 \mu_{i} \cos\theta (1+I) - 4 \mu_{i} \gamma (\cos\beta +1) - 2 \mu_{i} \alpha (\cos\theta + 1))\\
&  -D(4 \mu_{i} \sin\beta \gamma + 2\mu_{i} \sin\theta \alpha) + D(8\mu_{i} \sin\beta  + 4\mu_{i} \sin\theta)
\end{aligned}
\end{equation}
\begin{equation}
\begin{aligned}
B&= J( 4 \mu_{i} \gamma (\cos\beta -1) + 2 \mu_{i} \alpha (\cos\theta - 1))-D(4 \gamma \mu_{i} \sin\beta + 2\mu_{i} \alpha \sin\theta)
\end{aligned}
\end{equation}
where the sum on the two NN on the $x$ axis (see Fig. \ref{GSDJH0}b) is
\begin{equation}
\sum_{l} e^{-i\mathbf k\cdot (\mathbf R_{i}-\mathbf R_{l})}= 2\cos(k_x)\equiv 2  \alpha
\end{equation}
and the sum on the four NN on the oblique directions of the hexagon (see Fig. \ref{GSDJH0}b) is
\begin{equation}
\sum_{l} e^{-i \mathbf k \cdot (\mathbf R_{i}-\mathbf R_{l})}  = 4 \cos(k_x/2) \ \cos(\sqrt{3} k_y/2)\equiv 4\gamma
\end{equation}
Solving Eq. (\ref{secular}) for each given ($k_x,k_y$) one obtains the spin-wave frequency $\omega (k_x,ky)$:
\begin{equation}\label{SWspectrum}
(\hbar \omega)^2=A^2-B^2\ \ \rightarrow \hbar \omega=\pm \sqrt{A^2-B^2}
\end{equation}
Plotting $\omega (k_x,ky)$ in the space $(k_x,ky)$ one obtains the full spin-wave spectrum.

The spin length $\langle S^{z}_i\rangle$ (for all $i$, by symmetry) is given by (see technical details in Ref. \cite{DiepTM}):
\begin{equation}\label{lm2}
\langle S^{z}\rangle\equiv \langle S^{z}_i\rangle=\frac{1}{2}-
   \frac{1}{\Delta}
   \int
   \int dk_xdk_z
   \sum_{i=1}^{2}\frac{Q(E_i)}
   {\mbox{e}^{E_i/k_BT}-1}
\end{equation}
where $E_i (i=1,2)=\pm \sqrt{A^2-B^2}$ are the two solutions given above, and $Q(E_i)$ is the determinant (cofactor) obtained by replacing the first column of
$\mathbf M$ by $\mathbf C$ at $E_i$.

The spin length $\langle S^{z}\rangle$ at a given $T$ is calculated self-consistently by following the method given in Ref. \cite{DiepTM,Diep2017}.

\begin{figure}[h!]
\center
\vspace{-2cm}
\includegraphics[width= 12cm]{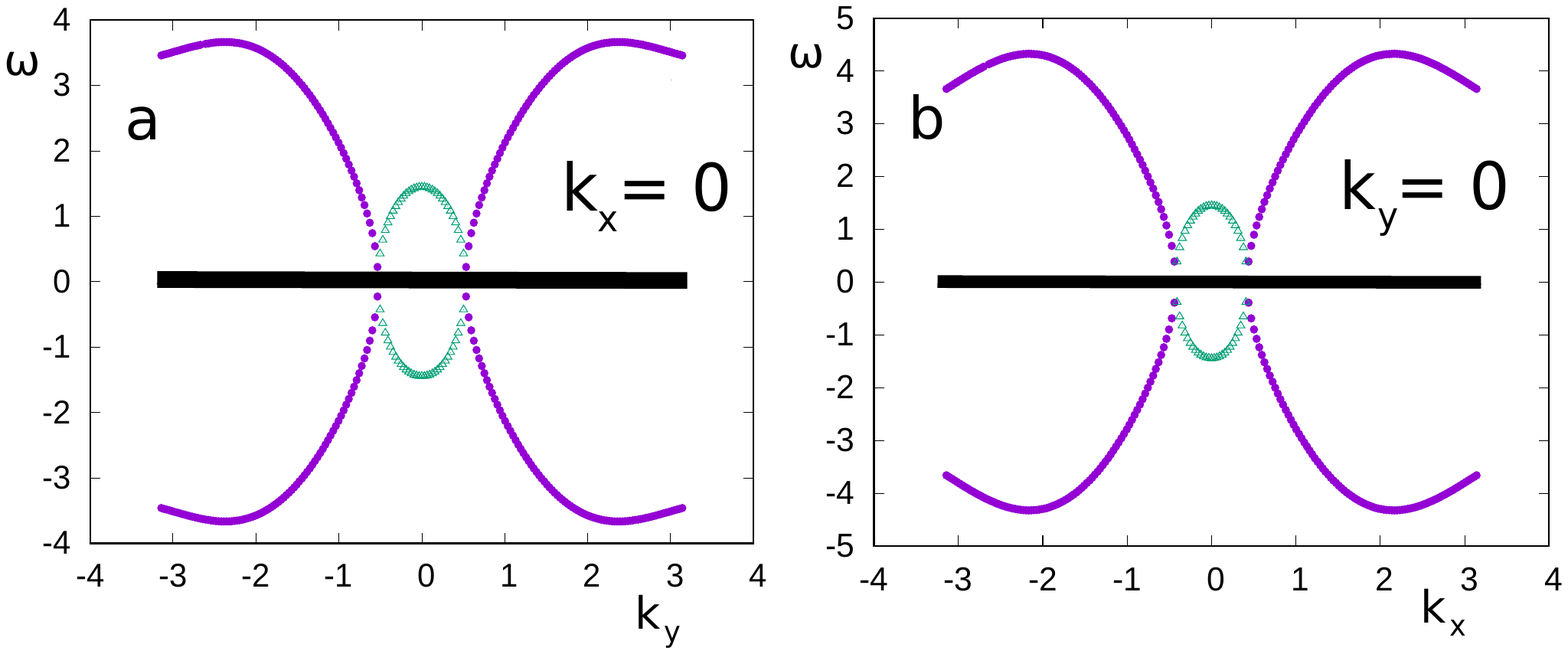}
\vspace{-7cm}
\caption{ (a) Spin-wave spectrum versus $k_y$ with $k_x=0$ at $T=0$ for $I= 0$, (b)
Spin-wave spectrum versus $k_x$ with $k_y=0$ at $T=0$ for $I= 0$. The magenta curves show the real frequency, while the green ones show the imaginary frequency. See text for comments. Parameters: $D = 0.5$, $J = -1$, $H=0$ where $\theta = 102$ degrees and  $\beta =  156$ degrees (see the spin configuration shown in Fig. \ref{GSDJH0}), $\hbar=1$.}\label{SWD1}
\end{figure}

Let us show the spin-wave spectrum $\omega$ (taking $\hbar=1$)  for the case of $J=-1$ and $D=0.5$ in Fig. \ref{SWD1} versus $k_y$ with $k_x=0$ (Fig. \ref{SWD1}a) and versus $k_x$ for $k_y=0$ (Fig. \ref{SWD1}b).  In order to see the effect of the DM interaction alone we take the anisotropy $I=0$.
One observes here that for a range of small wave-vectors the spin-wave frequency is imaginary.  The spin waves corresponding to these modes do not propagate in the system.  Why do we have this case here? The answer is that when the NN make a large angle (perpendicular NN, for example), one cannot define a wave vector in that direction. Physically, when $k$ is small the $B$ coefficient is larger than $A$ in Eq. (\ref{SWspectrum}) giving rise to imaginary $\omega$. Note that the anisotropy $I$ is contained in $A$ so that increasing $I$ for small $k$ will result in $A>B$ making $\omega$ real.

\begin{figure}[h!]
\center
\vspace{-3cm}
\includegraphics[width= 12cm]{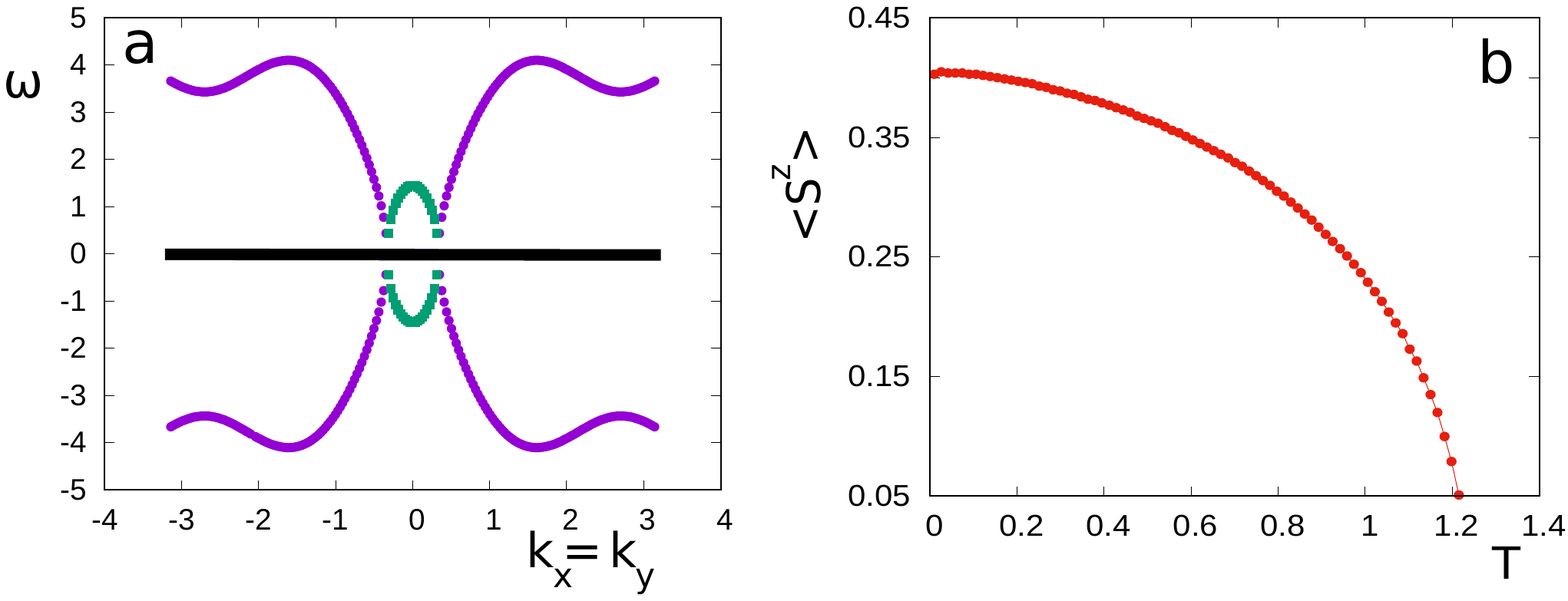}
\vspace{-7cm}
\caption{ (a) Spin-wave spectrum versus $k_x=k_y$ at $T=0$ for $I= 0$. The magenta curves show the real frequency, while the green ones show the imaginary frequency. See text for comments, (b) The spin length $S^z$ versus temperature $T$ ($k_B=1$). Parameters: $D = 0.5$, $J = -1$, $H=0$ where $\theta = 102$ degrees and  $\beta =  156$ degrees (see the spin configuration shown in Fig. \ref{GSDJH0}).}\label{SWD2}
\end{figure}

We show now in Fig. \ref{SWD2}a the spectrum along the axis $k_x=k_y$ at $T=0$ for $I= 0$. Again here the frequency is imaginary for small $k$, as in the previous figure.  The spin length $<S^z>$ along the local quantization axis is shown in Fig. \ref{SWD2}b. Several remarks are in order: i) At $T=0$, the spin length is not equal to $1/2$ as in ferromagnets because of the zero-point spin contraction due to antiferromagnetic interactions (see Ref. \cite{DiepTM}), its length is $\simeq 0.40$, quite small; ii) the magnetic ordering is destroyed at $T\simeq 1.2$.

To close the present section, we note that in the case of perpendicular $\mathbf D$ considered above, we did not observe skyrmion textures when applying a perpendicular magnetic field: all spin configurations are no more planar, making the calculation of the spin-wave spectrum more difficult. This problem is left for a future investigation.

\section{Ground State and Phase Transition in the case of an in-plane $\mathbf D$ in a Perpendicular Applied Field}\label{SC}

The GS in the case of in-plane $\mathbf D$ in a perpendicular applied field has been studied by MC simulations where it has been shown that for $D=0.5$, the skyrmion crystal phase exists for $2.4<H<6.2$ at low $T$ \cite{Rosales,Liu,Mohylna,Osorio}.  The purpose of this section is to show that the steepest-descent method at $T=0$ confirms this result at least at the phase center where $H=[3-4]$.   

\begin{figure}[h!]
\centering
\includegraphics[width=12cm]{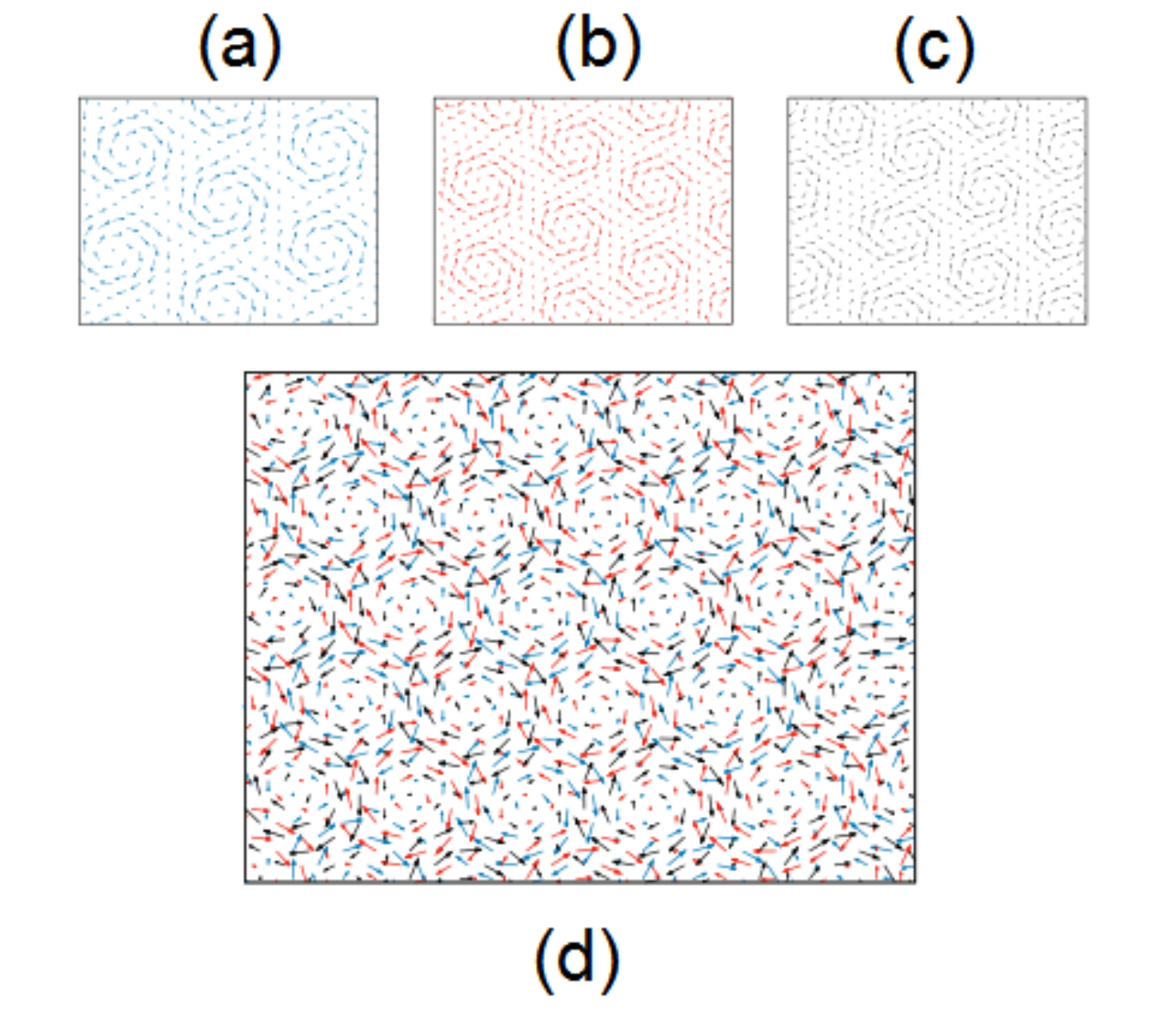}
\caption{In-plane $\mathbf D_{i,j}$:  The skyrmion crystal observed at $J$=-1, $D$=0.5 and $H$=3: A portion of skyrmion sublattice 1, 2 and 3 is respectively shown in (a), (b) and (c). Figure (d) shows the three interpenetrating antiferromagnetic skyrmion sublattices distinguished by three colors. }\label{ffig2}
\end{figure}

 We show in Fig. \ref{ffig2} the GS in the case of $J=-1$, $D=0.5$ and $H=3$ projected on the $xy$ plane where one observes a perfect crystal of skyrmions on a triangular lattice with a diameter of about ten spins.
It is interesting to note that the skyrmion crystal is composed of three interpenetrating skyrmion crystals represented in Fig. \ref{ffig2}a-c.  This finding is similar with the result obtained by Rosales et al. \cite{Rosales} by another method which is laborious (MC slow cooling).
The three-skyrmion-sublattice configuration has also found in the quantum version of our model using a mean-field MC  simulation \cite{Liu} and in a similar model including a single-ion anisotropy \cite{Mohylna} (we learn the results of this work after completing ours).  Our result together with these findings mean that the underlying frustrated antiferromagnetic triangular lattice naturally generates three-sublattice structure for an in-plane $\mathbf D_{i,j}$.  Note that in Fig. \ref{ffig2}, we see that the central spin in each sublattice points against the field direction ($S^z=-1$), and those in between the skyrmions are aligned with the field direction ($S^z=1$). In the between, the $xy$ components turn around the skyrmion center. In the three-sublattice presentation (Fig. \ref{ffig2}d), there is no central spin but rather the central triangle with three spins pointing against the field direction.

We note that the antiferromagnetic triangular lattice without the DM interaction can  also generate a skyrmion crystal under an applied perpendicular field if one includes the NNN and/or third NN interactions (see Ref. \cite{Okubo}).


%

 A good periodicity of the skyrmion phase is  found here in the case $H=3$.  Our method however did not give a perfect skyrmion far from the phase center. It may be due to the inefficiency of the method or to other mechanisms at $T=0$. One of these is the phenomenon called "order by disorder" in which there is an absence of a long-range order in the GS but the system selects an ordered phase as soon as $T$ becomes finite \cite{Diephcp}.  In the present case, we do not know the origin of the problem. This point is interesting and left for a future investigation.

%



In the following, we study the phase transition of the skyrmion crystal shown above when we increase the temperature $T$.

To see if the skyrmions and its crystalline structure are stable at finite temperatures, we use MC simulations to calculate the averaged energy per spin $\langle E \rangle$, the specific heat $C_v$ and the order parameter $Q$.
These quantities are defined as follows

\begin{eqnarray}
\langle E \rangle &=& \langle \mathcal{H} \rangle /(2N) \\
C_v&=&\frac{\langle E^2 \rangle -\langle E \rangle^2}{k_BT^2}\\
Q(T)&=&\frac{1}{N(t_a-t_0)}\sum_i |\sum_{t=t_0}^{t_a} \mathbf S_i (T,t)\cdot \mathbf S_i(T=0)|\label{OrderP}\\
\end{eqnarray}
where $\mathbf S_i (T,t)$ is the $i$-th spin at the time $t$, at temperature $T$, and $\mathbf S_i (T=0)$ is its state in the GS. $t_0$ is the starting time of the average and $t_a$ is the length of time for averaging.

Note that the order parameter is defined as the time average of the projection of the spin configuration of the entire system at a given temperature $T$ on its GS.  It has been introduced for the first time in our earlier work \cite{ElHog2018} in order to see the symmetry breaking of complicated non-collinear spin configurations. At very low $T$, the deviation of each spin with respect to its orientation in the GS is very small so that $Q(T)$ is close to 1. At high $T$, each spin has a random orientation so that $Q(T)$ is zero. There is a transition temperature $T_c$ at which $Q(T)\rightarrow 0$ with increasing $T$.

\begin{figure}[h!]
\centering
\includegraphics[width=12cm]{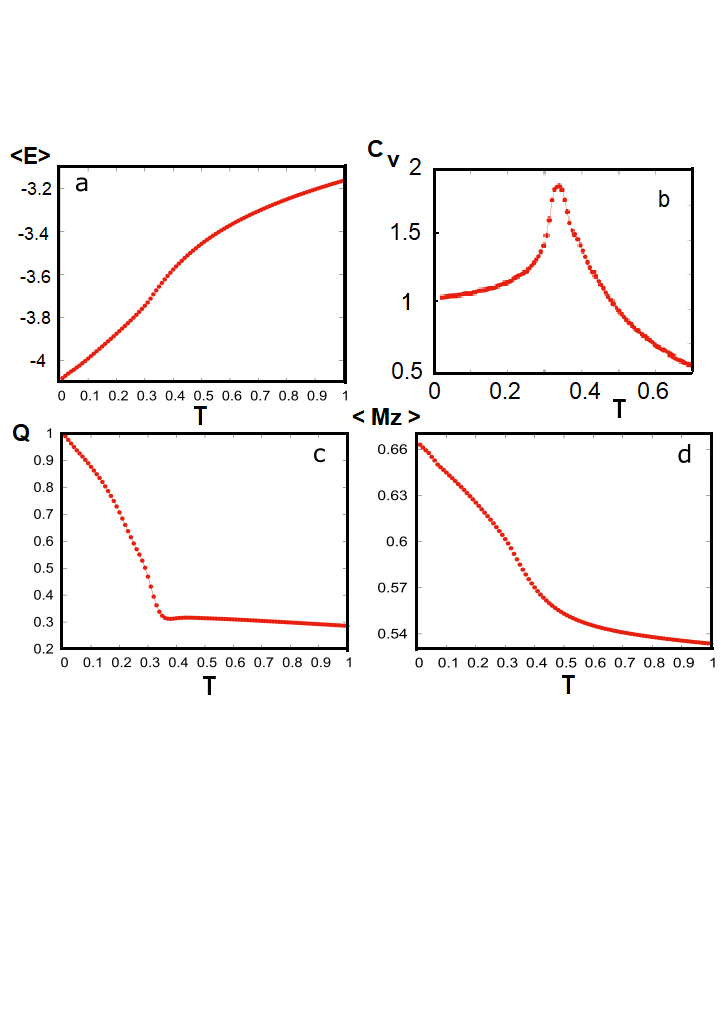}
\vspace{-5cm}
\caption{Results for $J$=-1, $D=0.5$, $H=3$: (a) Energy vs $T$, (b) Specific heat vs $T$, (c) Order parameter vs $T$, (d) $z$-component of the magnetization versus $T$. }\label{transit}
\end{figure}

We have performed MC simulations starting from the GS by a slow heating.  We used samples of sizes mostly $100\times 100$, but some calculations have been done with several sizes  up to $200\times 200$ to check the size dependence of the transition temperature. No significant variation was found. Let us show in Fig. \ref{transit} the energy, the specific heat, the order parameter and the absolute value of the $z$-component of the magnetization, as functions of temperature $T$. Several remarks are in order:

(i) the energy changes its curvature at $T_c=0.320\pm 0.005$ which corresponds to the peak of the specific heat.  Note that the value of our specific heat is in agreement with the Refs. \cite{Rosales,Mohylna}.

(ii) the order parameter $Q$ changes its curvature also at $T_c$ indicated above.  The fact that $Q$ is not zero for $T>T_c$ is due to the magnetic field which blocks the system in the $z$ direction. This is seen by the $z$-component of the magnetization shown in the last figure for $T>T_c$.

(iii) Note that the $z$-component of the magnetization is equal to about $2/3$ near $T=0$.

As said earlier, results for lattice sizes up to $200\times 200$ are not significantly different. Our conclusion is that the skyrmion  crystal observed in Fig. \ref{ffig2} is stable up to a finite temperature. The transition to the disordered phase (in the magnetic field) occurs at $T_c=0.320\pm 0.005$. Results for other values of $H$ yields the same conclusion but with $T_c$ slightly different.

\section{Conclusion}\label{Concl}

In this paper, in the absence of the magnetic field and the exchange interaction $J$,  we have shown  analytically for the first time that the ground state of the perpendicular  DM interaction on a triangular lattice is periodic, characterized by two angles.  When both $D$ and $J$ are present in zero field, the ground state is still periodic and characterized also by two angles, as we found numerically using the steepest-descent method. The periodicity of the ground state in this case has allowed us to use the Green's function method, devised for non-collinear spin configurations, to calculate the spin-wave spectrum and the local magnetization as a function of temperature. The results show that for a region of small $k$, spin waves cannot propagate in  the system. The physical meaning of this point has been discussed.

We have also checked that by the use of a very fast steepest-descent method at $T=0$ at the phase center, the frustrated antiferromagnetic Heisenberg triangular lattice including a Dzyaloshinskii-Moriya interaction with an in-plane DM vector  gives birth to a skyrmion crystal at $T=0$ under an applied perpendicular magnetic field, in agreement with previous studies using laborious MC methods to attain low, but finite $T$ configurations.  This skyrmion crystal, composed of three  antiferromagnetic skyrmion sublattices, is of the Bloch type. We have studied the phase transition of this skyrmion crystal by Monte Carlo simulations. Our order parameter $Q$ together with other physical quantities indicate that the skyrmion crystal is stable up to a rather high temperature.  This is very important because applications can be devised only at finite temperatures. Our steepest-descent method however did not give perfect skyrmion crystals at $T=0$ away from the phase center, unlike in the previous works. This point is left for a future study.

\acknowledgments{The author I.F.S. thanks for the support of the State assignment of Russian Federation for the implementation of scientific research by laboratories (Order MN-8/1356 of 09/20/2021).}

\end{document}